\documentclass[fleqn,twocolumn,showpacs,preprintnumbers,showkeys]{revtex4}
\usepackage{graphicx}
\usepackage{amssymb}
\usepackage{amsmath}
\usepackage{bm}
\usepackage[flushleft]{caption2}

\begin{document}
\setcounter{page}{1}
\title{Some theorems in gravitational and electromagnetic fields}
\author{Zihua Weng}
\email{xmuwzh@xmu.edu.cn.}
\affiliation{School of Physics and
Mechanical \& Electrical Engineering,
\\Xiamen University, Xiamen 361005, China}

\begin{abstract}
The paper discusses the influences of velocity curl and field
strength on some theorems in the electromagnetic field and
gravitational field. With the characteristics of the algebra of
quaternions, the theorem of linear momentum, conservation of linear
momentum, and conservation of angular momentum etc. can be deduced
from the quaternionic definitions of physical quantities. And the
strength of gravitational field and electromagnetic field have an
influence on some theorems directly. While the velocity curl has an
effect on some theorems also.
\end{abstract}



\maketitle

\section{INTRODUCTION}

The quaternions can be used to describe conservation laws and
theorems in the electromagnetic field \cite{maxwell} and the
gravitational field \cite{chu}, including the theorem of linear
momentum, the conservation of linear momentum, and the conservation
of angular momentum, etc. \cite{newton}

The concept of the linear momentum was originated by a number of
great scientists. The linear momentum of the particle and the
conservation of linear momentum were introduced by R. Descartes in
1644. Later, the concept of linear momentum extended from the
particle to the electromagnetic field etc. Further, this concept
covered the quantum mechanics \cite{dirac, adler}.

The angular momentum is an important concept in the physics, with
numerous applications. The concept of the angular momentum covers
that in the gravitational field and electromagnetic field.

With the features of the algebra of quaternions \cite{hamilton}, we
can obtain theorem of linear momentum, conservation of linear
momentum, conservation of angular momentum, and conservation of
electric current etc. in the gravitational field and electromagnetic
field. And we find that the velocity curl and strength have a few
influences on the theorems and conservation laws in the
gravitational field and electromagnetic field.

\section{Gravitational field}

The theorems in gravitational field can be described by quaternions.
In the quaternion space, the coordinates are $r_0$, $r_1$, $r_2$,
and $r_3$, with the basis vector $\mathbb{E}_g = (1 ,
\emph{\textbf{i}}_1 , \emph{\textbf{i}}_2 , \emph{\textbf{i}}_3)$.
Where, $r_0 = v_0 t$; $v_0$ is the speed of light beam, and $t$ is
the time. The radius vector $\mathbb{R} = r_0 + \Sigma (r_j
\emph{\textbf{i}}_j)$, and the velocity $\mathbb{V} = v_0 + \Sigma
(v_j \emph{\textbf{i}}_j)$, $j = 1, 2, 3$ ; $i = 0, 1, 2, 3 $ .

The gravitational potential is,
\begin{eqnarray}
\mathbb{A} = a_0 + \Sigma (a_j \emph{\textbf{i}}_j)
\end{eqnarray}
and the strength $\mathbb{B}$ of gravitational field.
\begin{eqnarray}
\mathbb{B} = \lozenge \circ \mathbb{A} = b_0 + \Sigma (b_j
\emph{\textbf{i}}_j)
\end{eqnarray}
where, the $\circ$ denotes the quaternion multiplication. $ \lozenge
= \partial_0 + \Sigma (\emph{\textbf{i}}_j
\partial_j) ; ~\partial_i = \partial/\partial r_i$;
$\emph{\textbf{a}} = \Sigma (a_j \emph{\textbf{i}}_j)$; $\nabla =
\Sigma (\emph{\textbf{i}}_j \partial_j)$.

The gravitational strength $\mathbb{B}$ includes two components,
$\textbf{g}/v_0 = \partial_0 \emph{\textbf{a}} + \nabla a_0 $ and
$\textbf{b} = \nabla \times \emph{\textbf{a}} $ , with the gauge
$b_0 = 0$.
\begin{eqnarray}
\textbf{g}/v_0 = && \emph{\textbf{i}}_1 ( \partial_0 a_1 +
\partial_1 a_0 ) + \emph{\textbf{i}}_2 ( \partial_0 a_2 + \partial_2
a_0 )
\nonumber\\
&& + \emph{\textbf{i}}_3 ( \partial_0 a_3 + \partial_3 a_0 )
\\
\textbf{b} = && \emph{\textbf{i}}_1 ( \partial_2 a_3 -
\partial_3 a_2 ) + \emph{\textbf{i}}_2 ( \partial_3 a_1 - \partial_1
a_3 )
\nonumber\\
&& + \emph{\textbf{i}}_3 ( \partial_1 a_2 - \partial_2 a_1 )
\end{eqnarray}
where, $\textbf{b} = 0$ in the Newtonian gravity.

The source $\mathbb{S}$ of gravitational field includes the linear
momentum density $\mathbb{S}_g = m \mathbb{V} $ and an extra part
$v_0 \triangle m$ .
\begin{eqnarray}
\mu \mathbb{S}  = - ( \mathbb{B}/v_0 + \lozenge)^* \circ \mathbb{B}
= \mu_g^g \mathbb{S}_g - \mathbb{B}^* \circ \mathbb{B}/v_0
\end{eqnarray}
where, $m$ is the mass density; $*$ denotes the quaternion
conjugate; $\mu$ and $\mu_g^g$ are the constants; $\mathbb{B}^*
\circ \mathbb{B}/(2\mu_g^g)$ is the field energy density; $\triangle
m = - \mathbb{B}^* \circ \mathbb{B}/(\mu_g^g v_0^2)$.

The applied force density $\mathbb{F}$ is defined from the linear
momentum density $\mathbb{P} = \mu \mathbb{S} / \mu_g^g$ . And the
latter is the extension of the $\mathbb{S}_g$.
\begin{eqnarray}
\mathbb{F} = v_0 (\mathbb{B}/v_0 + \lozenge )^* \circ \mathbb{P}
\end{eqnarray}
where, the applied force density includes the inertial force density
and the gravitational force density, etc.

The angular momentum density $\mathbb{L}$ is defined from the linear
momentum density $\mathbb{P} = \mu \mathbb{S} / \mu_g^g$ . The
latter is the extension of the $\mathbb{S}_g = m \mathbb{V} $ .
\begin{eqnarray}
\mathbb{L} = (\mathbb{R} + k_{rx} \mathbb{X} ) \circ \mathbb{P}
\end{eqnarray}
where, $\mathbb{X} = \Sigma (x_i \emph{\textbf{i}}_i)$, with
$k_{rx}$ being the coefficient.

And the total energy density $\mathbb{W}$ is defined from the
angular momentum density $\mathbb{L}$ .
\begin{eqnarray}
\mathbb{W}  = v_0 ( \mathbb{B}/v_0 + \lozenge) \circ \mathbb{L}
\end{eqnarray}
where, the total energy includes the potential energy, the kinetic
energy, the torque, and the work, etc. in the gravitational field.

By the total energy density $\mathbb{W}$, we obtain the external
power density $\mathbb{N}$ as follows.
\begin{eqnarray}
\mathbb{N}  = v_0 ( \mathbb{B}/v_0 + \lozenge)^* \circ \mathbb{W}
\end{eqnarray}
where, the external power density $\mathbb{N}$ includes the power
density etc. in the gravitational field.

\begin{table}[t]
\caption{\label{tab:table1}The quaternion multiplication table.}
\begin{ruledtabular}
\begin{tabular}{ccccc}
$ $ & $1$ & $\emph{\textbf{i}}_1$  & $\emph{\textbf{i}}_2$ &
$\emph{\textbf{i}}_3$  \\
\hline $1$ & $1$ & $\emph{\textbf{i}}_1$  & $\emph{\textbf{i}}_2$ &
$\emph{\textbf{i}}_3$  \\
$\emph{\textbf{i}}_1$ & $\emph{\textbf{i}}_1$ & $-1$ &
$\emph{\textbf{i}}_3$  & $-\emph{\textbf{i}}_2$ \\
$\emph{\textbf{i}}_2$ & $\emph{\textbf{i}}_2$ &
$-\emph{\textbf{i}}_3$ & $-1$ & $\emph{\textbf{i}}_1$ \\
$\emph{\textbf{i}}_3$ & $\emph{\textbf{i}}_3$ &
$\emph{\textbf{i}}_2$ & $-\emph{\textbf{i}}_1$ & $-1$
\end{tabular}
\end{ruledtabular}
\end{table}

\subsection{Theorem of linear momentum}

In the quaternion space, the inertial mass density is $m$, and the
gravitational mass density $\widehat{m} = m + \triangle m$. The
linear momentum density $\mathbb{P} = p_0 + \Sigma (p_j
\emph{\textbf{i}}_j )$ , with $p_0 = \widehat{m} v_0$ and $p_j = m
v_j $ .

By Eq.(6), the applied force density $\mathbb{F}$ is
\begin{eqnarray}
\mathbb{F} = f_0 + \Sigma (f_j \emph{\textbf{i}}_j )
\end{eqnarray}
where, $f_0 = \partial p_0 / \partial t + v_0 \Sigma (  \partial p_j
/ \partial r_j ) + \Sigma ( b_j p_j ) $ .

In the quaternion space, the vectorial part $\emph{\textbf{f}}$ of
applied force density $\mathbb{F}$ can be decomposed from Eq.(10).
\begin{eqnarray}
\emph{\textbf{f}} = && \Sigma (f_j \emph{\textbf{i}}_j )
\nonumber\\
= && v_0 \partial_0 \emph{\textbf{p}} + p_0 \emph{\textbf{h}}^*
\nonumber\\
&&  + \emph{\textbf{h}}^* \times \emph{\textbf{p}} + v_0 \nabla^*
p_0 + v_0 \nabla^* \times \emph{\textbf{p}}
\end{eqnarray}
where, $\emph{\textbf{h}} = \Sigma (b_j \emph{\textbf{i}}_j) $,
$\emph{\textbf{p}} = \Sigma (p_j \emph{\textbf{i}}_j) $, $\nabla =
\Sigma ( \emph{\textbf{i}}_j \partial_j ) $.

The above can be rewritten as follows,
\begin{eqnarray}
\emph{\textbf{f}} = \partial \emph{\textbf{p}} / \partial t +
\emph{\textbf{F}}
\end{eqnarray}
where, $\emph{\textbf{F}} = p_0 \emph{\textbf{h}}^* + v_0 \nabla^*
p_0 + (\emph{\textbf{h}} + v_0 \nabla)^* \times \emph{\textbf{p}}$.
In some cases, there exist the $\emph{\textbf{f}} = 0$ .

In case of the time $t$ is only the independent variable, the
$\partial \emph{\textbf{p}} / \partial t$ will become the $d
\emph{\textbf{p}} / d t$. And then we obtain the theorem of linear
momentum.
\begin{eqnarray}
(\emph{\textbf{f}} - \emph{\textbf{F}}) dt = d ( m
\emph{\textbf{v}})
\end{eqnarray}
where, $\emph{\textbf{v}} = \Sigma (v_j \emph{\textbf{i}}_j)$ .

Further, if the $(\emph{\textbf{f}} - \emph{\textbf{F}}) = 0$, the
conservation of linear momentum can be derived from the above.
\begin{eqnarray}
d ( m \emph{\textbf{v}}) = 0
\end{eqnarray}

The above means the force density $(\emph{\textbf{f}} -
\emph{\textbf{F}})$ covers the gravity density etc., but does not
include the inertial force density from Eq.(11). And then the
theorem of linear momentum is only one of simple cases of Eq.(10).
So is the conservation of linear momentum.

\subsection{Theorem of angular momentum}
In the quaternion space, the vectorial part $\emph{\textbf{w}}$ of
total energy density $\mathbb{W}$ can be decomposed from Eq.(8).
\begin{eqnarray}
\emph{\textbf{w}} = && \Sigma (w_j \emph{\textbf{i}}_j )
\nonumber\\
= && v_0 \partial_0 \emph{\textbf{l}} + (\emph{\textbf{h}} + v_0
\nabla) l_0 + ( \emph{\textbf{h}} + v_0 \nabla) \times
\emph{\textbf{l}}
\end{eqnarray}

The above can be rewritten as follows,
\begin{eqnarray}
\emph{\textbf{w}} = \partial \emph{\textbf{l}} / \partial t +
\emph{\textbf{W}}
\end{eqnarray}
where, $\emph{\textbf{l}} = \Sigma {(m v_j) \emph{\textbf{i}}_j }$;
$\emph{\textbf{W}} = (\emph{\textbf{h}} + v_0 \nabla) l_0 + (
\emph{\textbf{h}} + v_0 \nabla) \times \emph{\textbf{l}}$; In some
cases, there exist the $\emph{\textbf{w}} = 0$ .

When the time $t$ is only the independent variable, the $\partial
\emph{\textbf{l}} / \partial t$ will become the $d \emph{\textbf{l}}
/ d t$ . And we obtain the theorem of angular momentum.
\begin{eqnarray}
(\emph{\textbf{w}} - \emph{\textbf{W}}) dt = d \emph{\textbf{l}}
\end{eqnarray}

In case of the $(\emph{\textbf{w}} - \emph{\textbf{W}}) = 0$, the
conservation of angular momentum can be derived from the above.
\begin{eqnarray}
d \emph{\textbf{l}} = 0
\end{eqnarray}

The above means the torque density $(\emph{\textbf{w}} -
\emph{\textbf{W}})$ covers the term $\emph{\textbf{h}} \times
\emph{\textbf{l}}$ etc., but does not include the term $\partial
\emph{\textbf{l}} / \partial t$ from Eq.(16). The theorem of angular
momentum is only one of simple cases of Eq.(15). And so is the
conservation of angular momentum in the gravitational field.

\subsection{Theorem of torque}
In the quaternion space, the vectorial part $\emph{\textbf{n}}$ of
external power density $\mathbb{N}$ can be decomposed from Eq.(9).
\begin{eqnarray}
\emph{\textbf{n}} = && \Sigma (n_j \emph{\textbf{i}}_j )
\nonumber\\
= && v_0 \partial_0 \emph{\textbf{w}} + (\emph{\textbf{h}} + v_0
\nabla)^* w_0
\nonumber\\
&& + ( \emph{\textbf{h}} + v_0 \nabla)^* \times \emph{\textbf{w}}
\end{eqnarray}

The above can be rewritten as follows,
\begin{eqnarray}
\emph{\textbf{n}} = \partial \emph{\textbf{w}} / \partial t +
\emph{\textbf{N}}
\end{eqnarray}
where, $\emph{\textbf{N}} = (\emph{\textbf{h}} + v_0 \nabla)^* w_0 +
( \emph{\textbf{h}} + v_0 \nabla)^* \times \emph{\textbf{w}}$ . In
some cases, there exist the $\emph{\textbf{n}} = 0$ .

When the time $t$ is only the independent variable, the $\partial
\emph{\textbf{w}} / \partial t$ will become $d \emph{\textbf{w}} / d
t$ . And we obtain the theorem of torque in the gravitational field.
\begin{eqnarray}
(\emph{\textbf{n}} - \emph{\textbf{N}}) dt = d \emph{\textbf{w}}
\end{eqnarray}

In case of the $(\emph{\textbf{n}} - \emph{\textbf{N}}) = 0$, the
conservation of torque can be derived from the above.
\begin{eqnarray}
d \emph{\textbf{w}} = 0
\end{eqnarray}

The above means that the vectorial part $(\emph{\textbf{n}} -
\emph{\textbf{N}})$ covers the term $\emph{\textbf{h}}^* \times
\emph{\textbf{w}}$ etc., but does not include the term $\partial
\emph{\textbf{w}} / \partial t$ from Eq.(20). The theorem of torque
is only one of simple cases of Eq.(19). And so is the conservation
of torque in the gravitational field.

\section{Gravitational and electromagnetic fields}

The gravitational field and electromagnetic field both can be
illustrated by the quaternion, and their quaternion spaces will be
combined together to become the octonion space. In other words, the
characteristics of gravitational field and electromagnetic field can
be described with the octonion space at the same time.

In the quaternion space for the gravitational field, the basis
vector $\mathbb{E}_g$ = ($1$, $\emph{\textbf{i}}_1$,
$\emph{\textbf{i}}_2$, $\emph{\textbf{i}}_3$), and the radius vector
$\mathbb{R}_g$ = ($r_0$, $r_1$, $r_2$, $r_3$), with the velocity
$\mathbb{V}_g$ = ($v_0$, $v_1$, $v_2$, $v_3$). For the
electromagnetic field, the basis vector $\mathbb{E}_e$ =
($\emph{\textbf{I}}_0$, $\emph{\textbf{I}}_1$,
$\emph{\textbf{I}}_2$, $\emph{\textbf{I}}_3$), the radius vector
$\mathbb{R}_e$ = ($R_0$, $R_1$, $R_2$, $R_3$), and the velocity
$\mathbb{V}_e$ = ($V_0$, $V_1$, $V_2$, $V_3$), with $\mathbb{E}_e$ =
$\mathbb{E}_g$ $\circ$ $\emph{\textbf{I}}_0$ .

The $\mathbb{E}_e$ is independent of the $\mathbb{E}_g$ . Both of
them can be combined together to become the basis vector
$\mathbb{E}$ of the octonion space.
\begin{eqnarray}
\mathbb{E} = (1, \emph{\textbf{i}}_1, \emph{\textbf{i}}_2,
\emph{\textbf{i}}_3, \emph{\textbf{I}}_0, \emph{\textbf{I}}_1,
\emph{\textbf{I}}_2, \emph{\textbf{I}}_3)
\end{eqnarray}

The radius vector $\mathbb{R} (r_0 , r_1 , r_2 , r_3 , R_0 , R_1 ,
R_2 , R_3 )$ in the octonion space is
\begin{eqnarray}
\mathbb{R} = r_0 + \Sigma (\emph{\textbf{i}}_j r_j ) + \Sigma
(\emph{\textbf{I}}_i R_i )
\end{eqnarray}
and the velocity $\mathbb{V} (v_0 , v_1 , v_2 , v_3 , V_0 , V_1 ,
V_2 , V_3 )$  is
\begin{eqnarray}
\mathbb{V} = v_0 + \Sigma (\emph{\textbf{i}}_j v_j ) + \Sigma
(\emph{\textbf{I}}_i V_i )
\end{eqnarray}
where, $r_0 = v_0 t$ ; $v_0$ is the speed of light; $t$ is the time;
The symbol $\circ$ denotes the octonion multiplication.

When the electric charge is combined with the mass to become the
electron or the proton etc., we obtain the $R_i \emph{\textbf{I}}_i
= r_i \emph{\textbf{i}}_i \circ \emph{\textbf{I}}_0$ and $ V_i
\emph{\textbf{I}}_i = v_i \emph{\textbf{i}}_i \circ
\emph{\textbf{I}}_0$ , with $\emph{\textbf{i}}_0 = 1$ .

The potential of the gravitational and electromagnetic fields are
$\mathbb{A}_g = (a_0 , a_1 , a_2 , a_3)$ and $\mathbb{A}_e = (A_0 ,
A_1 , A_2 , A_3)$ respectively. They are combined together to become
the potential $\mathbb{A} = \mathbb{A}_g + k_{eg} \mathbb{A}_e $,
with $k_{eg}$ being the coefficient.

The strength $\mathbb{B}(b_0, b_1, b_2, b_3, B_0, B_1, B_2, B_3)$
consists of the gravitational strength $\mathbb{B}_g$ and the
electromagnetic strength $\mathbb{B}_e$ . The gauge, $b_0 = 0$, and
$B_0 = 0$ .
\begin{eqnarray}
\mathbb{B} = \lozenge \circ \mathbb{A} = \mathbb{B}_g + k_{eg}
\mathbb{B}_e
\end{eqnarray}

The gravitational strength $\mathbb{B}_g$ in Eq.(2) includes two
components, $\textbf{g} = ( g_{01} , g_{02} , g_{03} ) $ and
$\textbf{b} = ( g_{23} , g_{31} , g_{12} )$, while the
electromagnetic strength $\mathbb{B}_e$ involves two parts,
$\textbf{E} = ( B_{01} , B_{02} , B_{03} ) $ and $\textbf{B} = (
B_{23} , B_{31} , B_{12} )$ .
\begin{eqnarray}
\textbf{E}/v_0 = && \emph{\textbf{I}}_1 ( \partial_0 A_1 +
\partial_1 A_0 ) + \emph{\textbf{I}}_2 ( \partial_0 A_2 + \partial_2
A_0 )
\nonumber\\
&& + \emph{\textbf{I}}_3 ( \partial_0 A_3 + \partial_3 A_0 )
\\
\textbf{B} = && \emph{\textbf{I}}_1 ( \partial_3 A_2 - \partial_2
A_3 ) + \emph{\textbf{I}}_2 ( \partial_1 A_3 - \partial_3 A_1 )
\nonumber\\
&& + \emph{\textbf{I}}_3 ( \partial_2 A_1 - \partial_1 A_2 )
\end{eqnarray}

In the octonion space, the electric current density $\mathbb{S}_e =
q \mathbb{V}_g \circ \emph{\textbf{I}}_0$ is the source for the
electromagnetic field, and the linear momentum density $\mathbb{S}_g
= m \mathbb{V}_g $ for the gravitational field. The source
$\mathbb{S}$ satisfies,
\begin{eqnarray}
\mu \mathbb{S} && = - ( \mathbb{B}/v_0 + \lozenge)^* \circ
\mathbb{B}
\nonumber\\
&& = \mu_g^g \mathbb{S}_g + k_{eg} \mu_e^g \mathbb{S}_e -
\mathbb{B}^* \circ \mathbb{B}/v_0
\end{eqnarray}
where, $k_{eg}^2 = \mu_g^g /\mu_e^g$; $q$ is the electric charge
density; $\mu_e^g$ is the constant; $*$ denotes the conjugate of
octonion.
\begin{eqnarray}
\mathbb{B}^* \circ \mathbb{B}/ \mu_g^g = \mathbb{B}_g^* \circ
\mathbb{B}_g / \mu_g^g + \mathbb{B}_e^* \circ \mathbb{B}_e / \mu_e^g
\end{eqnarray}

The force density $\mathbb{F}$ is defined from linear momentum
density $\mathbb{P} = \mu \mathbb{S} / \mu_g^g$ , which is the
extension of the $\mathbb{S}_g$ .
\begin{eqnarray}
\mathbb{F} = v_0 (\mathbb{B}/v_0 + \lozenge )^* \circ \mathbb{P}
\end{eqnarray}
where, the force density includes the gravity density, the inertial
force density, Lorentz force density \cite{lorentz}, and the
interacting force density between the electromagnetic strength with
magnetic moment, etc.

In the octonion space, the angular momentum density
\begin{eqnarray}
\mathbb{L} = (\mathbb{R} + k_{rx} \mathbb{X} ) \circ \mathbb{P}
\end{eqnarray}
where, $\mathbb{P} = \mu \mathbb{S} / \mu_g^g$ ; $\mathbb{X} =
\Sigma ( x_i \emph{\textbf{i}}_i ) + \Sigma ( X_i
\emph{\textbf{I}}_i )$ .

And the total energy density $\mathbb{W}$ is defined from the
angular momentum density $\mathbb{L}$ .
\begin{eqnarray}
\mathbb{W}  = v_0 ( \mathbb{B}/v_0 + \lozenge) \circ \mathbb{L}
\end{eqnarray}
where, the total energy includes the potential energy, the kinetic
energy, torque, and work, etc. in the gravitational field and the
electromagnetic field.

By means of the total energy density $\mathbb{W}$, we obtain the
external power density
\begin{eqnarray}
\mathbb{N}  = v_0 ( \mathbb{B}/v_0 + \lozenge)^* \circ \mathbb{W}
\end{eqnarray}
where, the external power density includes the power density in the
gravitational and electromagnetic fields.

\begin{table}[t]
\caption{\label{tab:table1}The octonion multiplication table.}
\begin{ruledtabular}
\begin{tabular}{ccccccccc}
$ $ & $1$ & $\emph{\textbf{i}}_1$  & $\emph{\textbf{i}}_2$ &
$\emph{\textbf{i}}_3$  & $\emph{\textbf{I}}_0$  &
$\emph{\textbf{I}}_1$
& $\emph{\textbf{I}}_2$  & $\emph{\textbf{I}}_3$  \\
\hline $1$ & $1$ & $\emph{\textbf{i}}_1$  & $\emph{\textbf{i}}_2$ &
$\emph{\textbf{i}}_3$  & $\emph{\textbf{I}}_0$  &
$\emph{\textbf{I}}_1$
& $\emph{\textbf{I}}_2$  & $\emph{\textbf{I}}_3$  \\
$\emph{\textbf{i}}_1$ & $\emph{\textbf{i}}_1$ & $-1$ &
$\emph{\textbf{i}}_3$  & $-\emph{\textbf{i}}_2$ &
$\emph{\textbf{I}}_1$
& $-\emph{\textbf{I}}_0$ & $-\emph{\textbf{I}}_3$ & $\emph{\textbf{I}}_2$  \\
$\emph{\textbf{i}}_2$ & $\emph{\textbf{i}}_2$ &
$-\emph{\textbf{i}}_3$ & $-1$ & $\emph{\textbf{i}}_1$  &
$\emph{\textbf{I}}_2$  & $\emph{\textbf{I}}_3$
& $-\emph{\textbf{I}}_0$ & $-\emph{\textbf{I}}_1$ \\
$\emph{\textbf{i}}_3$ & $\emph{\textbf{i}}_3$ &
$\emph{\textbf{i}}_2$ & $-\emph{\textbf{i}}_1$ & $-1$ &
$\emph{\textbf{I}}_3$  & $-\emph{\textbf{I}}_2$
& $\emph{\textbf{I}}_1$  & $-\emph{\textbf{I}}_0$ \\
\hline $\emph{\textbf{I}}_0$ & $\emph{\textbf{I}}_0$ &
$-\emph{\textbf{I}}_1$ & $-\emph{\textbf{I}}_2$ &
$-\emph{\textbf{I}}_3$ & $-1$ & $\emph{\textbf{i}}_1$
& $\emph{\textbf{i}}_2$  & $\emph{\textbf{i}}_3$  \\
$\emph{\textbf{I}}_1$ & $\emph{\textbf{I}}_1$ &
$\emph{\textbf{I}}_0$ & $-\emph{\textbf{I}}_3$ &
$\emph{\textbf{I}}_2$  & $-\emph{\textbf{i}}_1$
& $-1$ & $-\emph{\textbf{i}}_3$ & $\emph{\textbf{i}}_2$  \\
$\emph{\textbf{I}}_2$ & $\emph{\textbf{I}}_2$ &
$\emph{\textbf{I}}_3$ & $\emph{\textbf{I}}_0$  &
$-\emph{\textbf{I}}_1$ & $-\emph{\textbf{i}}_2$
& $\emph{\textbf{i}}_3$  & $-1$ & $-\emph{\textbf{i}}_1$ \\
$\emph{\textbf{I}}_3$ & $\emph{\textbf{I}}_3$ &
$-\emph{\textbf{I}}_2$ & $\emph{\textbf{I}}_1$  &
$\emph{\textbf{I}}_0$  & $-\emph{\textbf{i}}_3$
& $-\emph{\textbf{i}}_2$ & $\emph{\textbf{i}}_1$  & $-1$ \\
\end{tabular}
\end{ruledtabular}
\end{table}

\subsection{Theorem of linear momentum}
In the octonion space, the gravitational mass density $\widehat{m} =
m + \triangle m$, with $\triangle m = - \mathbb{B}^* \circ
\mathbb{B} / (\mu_g^g v_0^2)$. The linear momentum density
$\mathbb{P} = p_0 + \Sigma (p_j \emph{\textbf{i}}_j ) + \Sigma (P_i
\emph{\textbf{I}}_i ) $. And, $m$ is the inertial mass density; $P_i
= M V_i $; $M = k_{eg} \mu_e^g q / \mu_g^g$; $p_0 = \widehat{m}
v_0$; $p_j = m v_j $ .

By Eq.(31), the applied force density $\mathbb{F}$ is
\begin{eqnarray}
\mathbb{F} = f_0 + \Sigma (f_j \emph{\textbf{i}}_j ) + \Sigma (F_i
\emph{\textbf{I}}_i )
\end{eqnarray}

In the octonion space, the vectorial part $\emph{\textbf{f}}$ of
applied force density $\mathbb{F}$ can be decomposed from Eq.(35).
\begin{eqnarray}
\emph{\textbf{f}} = \Sigma (f_j \emph{\textbf{i}}_j ) + \Sigma (F_i
\emph{\textbf{I}}_i )
\end{eqnarray}

The above can be rewritten as follows,
\begin{eqnarray}
\emph{\textbf{f}} = \partial \emph{\textbf{z}} / \partial t +
\emph{\textbf{F}}
\end{eqnarray}
where, $\emph{\textbf{F}} = \left\{ \Sigma (f_j \emph{\textbf{i}}_j
) - \partial \emph{\textbf{z}} / \partial t \right\} + \Sigma (F_i
\emph{\textbf{I}}_i )$ , which is the extension of Eq.(12).
$\emph{\textbf{z}} = \Sigma {(m v_j) \emph{\textbf{i}}_j} $ .

When the time $t$ is only the independent variable, the $\partial
\emph{\textbf{z}} / \partial t$ will become $d \emph{\textbf{z}} / d
t$. And we obtain the theorem of linear momentum in the octonion
space.
\begin{eqnarray}
(\emph{\textbf{f}} - \emph{\textbf{F}}) dt = d \emph{\textbf{z}}
\end{eqnarray}

Further, if the $(\emph{\textbf{f}} - \emph{\textbf{F}}) = 0$, the
conservation of linear momentum can be derived from the above.
\begin{eqnarray}
d \emph{\textbf{z}} = 0
\end{eqnarray}

The above means the $(\emph{\textbf{f}} - \emph{\textbf{F}})$ covers
the gravity density and Lorentz force density etc., but does not
include the inertial force density from Eq.(37). And the theorem of
the linear momentum is only one of simple cases of Eq.(36). So is
the conservation of linear momentum in the gravitational field and
electromagnetic field.

\subsection{Theorem of angular momentum}

In the octonion space, from the angular momentum density,
\begin{eqnarray}
\mathbb{L} = l_0 + \Sigma (l_j \emph{\textbf{i}}_j ) + \Sigma (L_i
\emph{\textbf{I}}_i )
\end{eqnarray}
we have the total energy density
\begin{eqnarray}
\mathbb{W} = w_0 + \Sigma (w_j \emph{\textbf{i}}_j ) + \Sigma (W_i
\emph{\textbf{I}}_i )~.
\end{eqnarray}

In the octonion space, the vectorial part $\emph{\textbf{w}}$ of
total energy density $\mathbb{W}$ can be decomposed from Eq.(41).
\begin{eqnarray}
\emph{\textbf{w}} = \Sigma (w_j \emph{\textbf{i}}_j ) + \Sigma (W_i
\emph{\textbf{I}}_i )
\end{eqnarray}

The above can be rewritten as follows.
\begin{eqnarray}
\emph{\textbf{w}} = \partial \emph{\textbf{j}} / \partial t +
\emph{\textbf{W}}
\end{eqnarray}
where, $\emph{\textbf{W}} = \left\{  \Sigma (w_j \emph{\textbf{i}}_j
) - \partial \emph{\textbf{j}} / \partial t \right\} + \Sigma (W_i
\emph{\textbf{I}}_i )$, which is the extension in Eq.(16).
$\emph{\textbf{h}} = \Sigma (b_j \emph{\textbf{i}}_j)$;
$\emph{\textbf{j}} = \Sigma (l_j \emph{\textbf{i}}_j)$.

When the time $t$ is only the independent variable, the $\partial
\emph{\textbf{j}} / \partial t$ will become the $d \emph{\textbf{j}}
/ d t$ . And we obtain the theorem of angular momentum in the
octonion space.
\begin{eqnarray}
(\emph{\textbf{w}} - \emph{\textbf{W}}) dt = d \emph{\textbf{j}}
\end{eqnarray}

In case of the $(\emph{\textbf{w}} - \emph{\textbf{W}}) = 0$, the
conservation of angular momentum can be derived from the above.
\begin{eqnarray}
d \emph{\textbf{j}} = 0
\end{eqnarray}

The above means that the torque density $(\emph{\textbf{w}} -
\emph{\textbf{W}})$ covers the terms $\emph{\textbf{h}} \times
\emph{\textbf{j}}$ etc., but does not include the term $\partial
\emph{\textbf{j}} / \partial t$ from Eq.(43). And the theorem of
angular momentum is only one of simple cases of Eq.(42). So is the
conservation of angular momentum in the case for the gravitational
and electromagnetic fields.

\subsection{Theorem of torque}

In the octonion space, from the total energy density $\mathbb{W}$,
we obtain the external power density
\begin{eqnarray}
\mathbb{N} = n_0 + \Sigma (n_j \emph{\textbf{i}}_j ) + \Sigma (N_i
\emph{\textbf{I}}_i )
\end{eqnarray}

In the octonion space, the vectorial part $\emph{\textbf{n}}$ of
external power density $\mathbb{N}$ can be decomposed from Eq.(46).
\begin{eqnarray}
\emph{\textbf{n}} = \Sigma (n_j \emph{\textbf{i}}_j ) + \Sigma (N_i
\emph{\textbf{I}}_i )
\end{eqnarray}

The above can be rewritten as follows,
\begin{eqnarray}
\emph{\textbf{n}} = \partial \emph{\textbf{y}} / \partial t +
\emph{\textbf{N}}
\end{eqnarray}
where, $\emph{\textbf{N}} = \left\{  \Sigma (n_j \emph{\textbf{i}}_j
) - \partial \emph{\textbf{y}} / \partial t \right\} + \Sigma (N_i
\emph{\textbf{I}}_i )$, which is the extension in Eq.(20).
$\emph{\textbf{y}} = \Sigma (w_j \emph{\textbf{i}}_j)$.

When the time $t$ is only the independent variable, the $\partial
\emph{\textbf{y}} / \partial t$ will become the $d \emph{\textbf{y}}
/ d t$ . And then, we obtain the theorem of torque in the octonion
space.
\begin{eqnarray}
(\emph{\textbf{n}} - \emph{\textbf{N}}) dt = d \emph{\textbf{y}}
\end{eqnarray}

In case of the $(\emph{\textbf{n}} - \emph{\textbf{N}}) = 0$, the
conservation of torque can be derived from the above.
\begin{eqnarray}
d \emph{\textbf{y}} = 0
\end{eqnarray}

The above means that the vectorial part $(\emph{\textbf{n}} -
\emph{\textbf{N}})$ covers the terms $\emph{\textbf{h}}^* \times
\emph{\textbf{y}}$ etc., but does not include the term $\partial
\emph{\textbf{y}} / \partial t$ from Eq.(47). And the theorem of
torque is only one of simple cases of Eq.(46). So is the
conservation of torque in the case for coexistence of gravitational
field and electromagnetic field.

\subsection{Theorem of electric current}

In the octonion space, a new physical quantity $\mathbb{F}_q$ can be
defined from Eq.(35).
\begin{eqnarray}
\mathbb{F}_q = \mathbb{F} \circ \emph{\textbf{I}}_0^* =  F_0 +
\Sigma (F_j \emph{\textbf{i}}_j ) - \Sigma (f_i \emph{\textbf{I}}_i
)
\end{eqnarray}

In the octonion space, the vectorial part $\emph{\textbf{f}}_q$ of
applied force density $\mathbb{F}_q$ can be decomposed from Eq.(51).
\begin{eqnarray}
\emph{\textbf{f}}_q = \Sigma (F_j \emph{\textbf{i}}_j ) - \Sigma
(f_i \emph{\textbf{I}}_i )
\end{eqnarray}

The above can be rewritten as follows.
\begin{eqnarray}
\emph{\textbf{f}}_q = \partial \emph{\textbf{Z}} / \partial t +
\emph{\textbf{F}}_q
\end{eqnarray}
where, $\emph{\textbf{Z}} = \Sigma \left\{ ( M V_j
)\emph{\textbf{i}}_j \right\}$; $\emph{\textbf{F}}_q = \left\{
\Sigma (F_j \emph{\textbf{i}}_j ) - \partial \emph{\textbf{Z}} /
\partial t \right\} - \Sigma (f_i \emph{\textbf{I}}_i )$; In some
cases, $\emph{\textbf{f}}_q = 0$ .

When the time $t$ is only the independent variable, the $\partial
\emph{\textbf{Z}} / \partial t$ will become $d \emph{\textbf{Z}} / d
t$. And we obtain the theorem of electric current in the octonion
space.
\begin{eqnarray}
(\emph{\textbf{f}}_q - \emph{\textbf{F}}_q) dt = d \emph{\textbf{Z}}
\end{eqnarray}

Further, if the $(\emph{\textbf{f}}_q - \emph{\textbf{F}}_q) = 0$,
the conservation of electric current can be derived from the above.
\begin{eqnarray}
d \emph{\textbf{Z}} = 0
\end{eqnarray}

The above means the $(\emph{\textbf{f}}_q - \emph{\textbf{F}}_q)$
covers the gravity density and Lorentz force density etc., but does
not include the inertial force density from Eq.(53). And the theorem
of the electric current is only one of simple cases of Eq.(52). So
is the conservation of electric current in the gravitational field
and electromagnetic field.

\subsection{Theorem of magnetic momentum}

In the octonion space, a new physical quantity $\mathbb{W}_q$ can be
defined from Eq.(41).
\begin{eqnarray}
\mathbb{W}_q = \mathbb{W} \circ \emph{\textbf{I}}_0^* = W_0 + \Sigma
(W_j \emph{\textbf{i}}_j ) - \Sigma (w_i \emph{\textbf{I}}_i )
\end{eqnarray}

In the octonion space, the vectorial part $\emph{\textbf{w}}_q$ of
total energy density $\mathbb{W}_q$ can be decomposed from Eq.(56).
\begin{eqnarray}
\emph{\textbf{w}}_q = \Sigma (W_j \emph{\textbf{i}}_j ) - \Sigma
(w_i \emph{\textbf{I}}_i )
\end{eqnarray}

The above can be rewritten as follows.
\begin{eqnarray}
\emph{\textbf{w}}_q = \partial \emph{\textbf{J}} / \partial t +
\emph{\textbf{W}}_q
\end{eqnarray}
where, $\emph{\textbf{W}}_q = \left\{ \Sigma (W_j
\emph{\textbf{i}}_j ) - \partial \emph{\textbf{W}} / \partial t
\right\} - \Sigma (w_i \emph{\textbf{I}}_i )$; $\emph{\textbf{J}} =
\Sigma ( L_j \emph{\textbf{i}}_j )$; $\emph{\textbf{H}} = \Sigma (
B_j \emph{\textbf{i}}_j )$; In some cases, $\emph{\textbf{w}}_q = 0$
.

When the time $t$ is only the independent variable, the $\partial
\emph{\textbf{J}} / \partial t$ will become the $d \emph{\textbf{J}}
/ d t$ . And we obtain the theorem of magnetic momentum in the
octonion space.
\begin{eqnarray}
(\emph{\textbf{w}}_q - \emph{\textbf{W}}_q) dt = d \emph{\textbf{J}}
\end{eqnarray}

In case of the $(\emph{\textbf{w}}_q - \emph{\textbf{W}}_q) = 0$,
the conservation of magneitc momentum can be derived from the above.
\begin{eqnarray}
d \emph{\textbf{J}} = 0
\end{eqnarray}

The above means that the torque density $(\emph{\textbf{w}}_q -
\emph{\textbf{W}}_q)$ covers the terms $\emph{\textbf{H}} \times
\emph{\textbf{J}}$ etc., but does not include the term $\partial
\emph{\textbf{J}} / \partial t$ from Eq.(57). And the theorem of
magnetic momentum is only one of simple cases of Eq.(56). So is the
conservation of magnetic momentum in the case for the gravitational
and electromagnetic fields.

\subsection{Theorem of torque-like}

In the octonion space, a new physical quantity $\mathbb{N}_q$ can be
defined from Eq.(46).
\begin{eqnarray}
\mathbb{N}_q = \mathbb{N} \circ \emph{\textbf{I}}_0^* = N_0 + \Sigma
(N_j \emph{\textbf{i}}_j ) - \Sigma (n_i \emph{\textbf{I}}_i )
\end{eqnarray}

In the octonion space, the vectorial part $\emph{\textbf{n}}_q$ of
external power density $\mathbb{N}_q$ can be decomposed from
Eq.(61).
\begin{eqnarray}
\emph{\textbf{n}}_q = \Sigma (N_j \emph{\textbf{i}}_j ) - \Sigma
(n_i \emph{\textbf{I}}_i )
\end{eqnarray}

The above can be rewritten as follows.
\begin{eqnarray}
\emph{\textbf{n}}_q = \partial \emph{\textbf{Y}} / \partial t +
\emph{\textbf{N}}_q
\end{eqnarray}
where, $\emph{\textbf{Y}} = \Sigma ( W_j \emph{\textbf{i}}_j )$;
$\emph{\textbf{N}}_q = \left\{ \Sigma (N_j \emph{\textbf{i}}_j ) -
\partial \emph{\textbf{Y}} / \partial t \right\} - \Sigma (n_i
\emph{\textbf{I}}_i )$; In some cases, $\emph{\textbf{n}}_q = 0$ .

When the time $t$ is only the independent variable, the $\partial
\emph{\textbf{W}}_q / \partial t$ will become the $d
\emph{\textbf{W}}_q / d t$ . And then, we obtain the theorem of
torque-like in the octonion space.
\begin{eqnarray}
(\emph{\textbf{n}}_q - \emph{\textbf{N}}_q) dt = d \emph{\textbf{Y}}
\end{eqnarray}

In case of the $(\emph{\textbf{n}}_q - \emph{\textbf{N}}_q) = 0$,
the conservation of torque-like can be derived from the above.
\begin{eqnarray}
d \emph{\textbf{Y}} = 0
\end{eqnarray}

The above means that the vectorial part $(\emph{\textbf{n}}_q -
\emph{\textbf{N}}_q)$ covers the terms $\emph{\textbf{H}}^* \times
\emph{\textbf{Y}}$ etc., but does not include the term $\partial
\emph{\textbf{Y}} / \partial t$ from Eq.(62). And the theorem of
torque-like is only one of simple cases of Eq.(61). So is the
conservation of torque-like in the case for coexistence of
gravitational field and electromagnetic field.

\section{Influences of velocity curl}

In the octonion compounding space for coexistence of gravitational
field and electromagnetic field, the radius vector $\mathbb{R}$ will
be extended to the $\mathbb{\bar{R}} = \mathbb{R} +
k_{rx}\mathbb{X}$ , although their basis vector $\mathbb{E}$ keeps
unchanged.

\subsection{Compounding space}
In the octonion compounding space, the basis vector remains the same
as that in Eq.(23), while the radius vector $\mathbb{R}$ will be
extended to $\mathbb{\bar{R}}$, with the octonion quantity
$\mathbb{X} = \Sigma (x_i \emph{\textbf{i}}_i ) + \Sigma (X_i
\emph{\textbf{I}}_i )$.

The radius vector $\mathbb{R}$ in Eq.(24) and the velocity
$\mathbb{V}$ in Eq.(25) will be extended to $\mathbb{\bar{R}}$ and
$\mathbb{\bar{V}}= \mathbb{V} + k_{rx}\mathbb{A}$ respectively.
Their components can be written as follows.
\begin{eqnarray}
r_i~ \rightarrow && ~\bar{r}_i = r_i + k_{rx} x_i~;
\\
R_i~ \rightarrow && ~\bar{R}_i = R_i + k_{rx} X_i~;
\\
v_i~ \rightarrow && ~\bar{v}_i = v_i + k_{rx} a_i~;
\\
V_i~ \rightarrow && ~\bar{V}_i = V_i + k_{rx} A_i~.
\end{eqnarray}

In a similar way, the potential $\mathbb{A}$ and the strength
$\mathbb{B}$ in Eq.(26) will be extended to $\mathbb{\bar{A}} =
\mathbb{A} + k_{rx}\mathbb{V}$ and $\mathbb{\bar{B}} = \mathbb{B} +
k_{rx}\mathbb{U}$ respectively.
\begin{eqnarray}
a_i~ \rightarrow && ~\bar{a}_i = a_i + K_{rx} v_i~;
\\
A_i~ \rightarrow && ~\bar{A}_i = A_i + K_{rx} V_i~;
\\
b_i~ \rightarrow && ~\bar{b}_i = b_i + K_{rx} u_i~;
\\
B_i~ \rightarrow && ~\bar{B}_i = B_i + K_{rx} U_i~.
\end{eqnarray}
where, $K_{rx} = 1/k_{rx}$, $k_{rx} = 1$ .

The velocity curl is
\begin{eqnarray}
\mathbb{U} = \lozenge \circ \mathbb{V} = \Sigma (u_i
\emph{\textbf{i}}_i ) + \Sigma (U_i \emph{\textbf{I}}_i )~.
\nonumber
\end{eqnarray}
where, $ \lozenge = \Sigma (\emph{\textbf{i}}_i
\partial_i) ; ~\partial_i = \partial/\partial \bar{r}_i$; in
general, $\bar{r}_i \approx r_i $ .

\subsection{Field equations}
In the octonion compounding space, the gauge equations extend into,
\begin{eqnarray}
\bar{b}_0 = b_0 + K_{rx} u_0 = 0~,~\bar{B}_0 = B_0 + K_{rx} U_0 =
0~. \nonumber
\end{eqnarray}
and then the source $\mathbb{S}$ in Eq.(29) will become
$\mathbb{\bar{S}}$.
\begin{eqnarray}
\mu \mathbb{\bar{S}} = \mu_g^g \mathbb{\bar{S}}_g + k_{eg} \mu_e^g
\mathbb{\bar{S}}_e - \mathbb{\bar{B}}^* \circ
\mathbb{\bar{B}}/\bar{v}_0
\end{eqnarray}
where, $\bar{v}_0 = v_0 + k_{rx} a_0$ , $\mathbb{\bar{S}}_g = m
(\mathbb{V}_g + k_{rx} \mathbb{A}_g)$ , $\mathbb{\bar{S}}_e = q
(\mathbb{V}_g \circ \emph{\textbf{I}}_0 + k_{rx} \mathbb{A}_e ) $ .

The applied force density $\mathbb{\bar{F}}$ is defined from the
linear momentum density $\mathbb{\bar{P}} = \mu \mathbb{\bar{S}} /
\mu_g^g$ .
\begin{eqnarray}
\mathbb{\bar{F}} = \bar{v}_0 (\mathbb{\bar{B}}/\bar{v}_0 + \lozenge
)^* \circ \mathbb{\bar{P}}
\end{eqnarray}

In the octonion space, the total energy density
$\mathbb{\bar{\bar{W}}}$ is defined from the angular momentum
density $\mathbb{\bar{L}} = \mathbb{\bar{R}} \circ
\mathbb{\bar{P}}$,
\begin{eqnarray}
\mathbb{\bar{W}}  = \bar{v}_0 ( \mathbb{\bar{B}}/\bar{v}_0 +
\lozenge) \circ \mathbb{\bar{L}}
\end{eqnarray}
and we obtain the external power density
\begin{eqnarray}
\mathbb{\bar{N}}  = \bar{v}_0 ( \mathbb{\bar{B}}/\bar{v}_0 +
\lozenge)^* \circ \mathbb{\bar{W}}~.
\end{eqnarray}

In the octonion compounding space, the above means that the velocity
curl has an influence on field equations. Similarly, the curl has an
impact on several theorems, including the theorem of linear
momentum, conservation of linear momentum, and theorem of angular
momentum etc. in the case for the gravitational field and
electromagnetic field.

\section{CONCLUSIONS}

In the quaternion spaces, from the definition of applied force, we
obtain the theorem of linear momentum and the conservation of linear
momentum. Similarly, we have the theorem of angular momentum and the
conservation of angular momentum etc. And we find the strength in
the gravitational field has an influence on the theorems and
conservation laws about the linear momentum.

In the octonion spaces, beside the above conservation laws and
theorems in gravitational field, we can obtain the conservation of
electric current and the conservation of magnetic moment etc. in the
electromagnetic field. And we find the strength in the
electromagnetic field has an influence on the theorems and
conservation laws about the electric current.

In the octonion compounding spaces, we can obtain the conservation
laws and theorems similarly, and find the velocity curl has an
influence on the theorems and conservation laws in the
electromagnetic field and the gravitational field.

It should be noted the study for the conservation laws and the
theorems of physical quantities examined only some simple cases
under the Galilean transformation in the electromagnetic field and
gravitational field. Despite its preliminary character, this study
can clearly indicate the conservation of linear momentum etc. are
only some of simple inferences due to low velocity curl and weak
strength of electromagnetic and gravitational fields. For the future
studies, the research will concentrate on only some predictions
about the conservation laws and the theorems with high velocity curl
and strong strength in the electromagnetic and gravitational fields.

\begin{acknowledgments}
This project was supported partially by the National Natural Science
Foundation of China under grant number 60677039.
\end{acknowledgments}

\end{document}